\documentclass[12pt]{article}
\newcommand{\1}{\begin{equation}}
\newcommand{\2}{\end{equation}}
\newcommand{\ea}{\begin{eqnarray}}
\newcommand{\ee}{\end{eqnarray}}
\newcommand{\bee}{\begin{eqnarray*}}
\newcommand{\eee}{\end{eqnarray*}}

\newcommand{\der}[1]{{\frac{\rm d}{{\rm d}#1}}}
\newcommand{\dnd}[2]{{\frac{{\rm d}#2}{{\rm d}#1}}}

\newcommand{\op}[1]{\hat{#1}}
\newcommand{\erw}[1]{\left\langle\, #1\,\right\rangle}

\newcommand{\de}{{\!\rm d}}
\newcommand{\e}{{\rm e}}

\newcommand{\g}{{\!\,=\,\!}}

\newcommand{\ii}{{\rm i}}

\newcommand{\sa}{\left[ \begin{array} {c} }
\newcommand{\se}{\end{array}\right]}

\usepackage{texdraw}
\usepackage{graphicx}
\begin{document}
\begin{center}
{\huge\bf Density excitations of a harmonically trapped ideal gas }\\*[8mm]
{\large Jai Carol Cruz\cite{cruz:09}, C. N. Kumar, K. N. Pathak}\\*[1.5mm]
{\small Department of Physics, Panjab University, Chandigarh, India\\*[3mm]
and}\\*[2.5mm]
{\large J. Bosse}\\*[1.5mm]
{\small Institute of Theoretical Physics, Freie Universit{\"a}t Berlin, Berlin, Germany\\*[2mm]
( $6\,{\rm th}$ March, 2009 )}\\
*[1cm]

{\large Abstract}
\end{center}

\begin{quote}
The dynamic structure factor $S({\bf q},\omega)$ of a harmonically trapped Bose gas has been calculated  well above the Bose--Einstein condensation temperature by treating the gas cloud as a canonical ensemble of noninteracting classical particles.  The static structure factor is found to vanish $\propto q^2$ in the long--wavelength limit. We also incorporate a relaxation mechanism phenomenologically by including a stochastic friction force to study $S({\bf q},\omega)$. A significant temperature dependence of the density--fluctuation spectra is found.
The Debye--Waller factor has been calculated for the trapped thermal cloud as function of $q$ and the number ${\cal N}$ of atoms. A substantial difference is found for small-- and large--${\cal N}$ clouds.
\end{quote}

\section{Introduction}

There has been much interest in the study of trapped atomic gases
during more than a decade.\cite{pes:02,pis:03}
Experimentalists have observed not only density profiles of these
systems but they have  successfully detected  particle--number and --density fluctuations in ultra--cold atomic gases.\cite{grs:05} Progress has been made in understanding spatial and temporal correlations of density fluctuations in a variety of inhomogeneous
cold--atom systems \cite{bbg:02, mkp:03, cac:06, bes:98} for $T\g0$
as well as below the Bose--Einstein condensation (BEC) temperature
$T_c$. In addition there have been extensive studies of dynamical
density fluctuations for homogeneous
systems. \cite{pin:66, ham:05}

The purpose of this work is to calculate the  diagonal elements of the matrix of time--dependent correlation functions of density fluctuations for
a harmonically trapped classical ideal gas, which determine the coherent scattering properties of the system. Motivation for calculating the dynamical structure factor is its expected strong temperature dependence  which could be of use in estimating the cloud temperature experimentally.
Moreover, the exactly solvable model not only provides the limiting case of a system in the
absence of interactions but also furnishes a useful result for trapped dilute
gases well above the BEC temperature. Justification for a classical
treatment is based on the fact that for an atom trap of typical trap
frequency $\Omega\g 120\times 2\pi\,{\rm s^{-1}}$ containing ${\cal
N}\g2\times10^4$ rubidium atoms which implies a critical temperature
$T_c\g\hbar\Omega {\cal N}^{1/3}/[\zeta(3)^{1/3}k_{\rm
B}]\approx147\,{\rm nK}$ of Bose--Einstein condensation \cite[Chap.
2.2]{pes:02}, the ratio $\hbar\Omega$ over $k_{\rm B}T$ is very
small in the temperature range well above $T_c$, where many
experiments are performed, because $\hbar\Omega/(k_{\rm
B}T_c)\approx 0.04$. Moreover, at these temperatures the reduced
density $\rho\g n\lambda^3\approx0.01$ is small implying
$\lambda\approx0.008\,L$, i.e. the deBroglie wavelength $\lambda$ is
still very small compared to the linear extension $L$ of the atom
cloud.

In addition to the restoring force acting on each particle due to the trap potential, we allow for a frictional force when studying the dynamical behavior of the trapped gas. Simultaneously, a time--dependent random force is assumed to act on each oscillator.
This stochastic force will keep particles moving, thus preventing them from being slowed down by friction and coming to a rest in the trap--potential minimum. Thermal averages are therefore understood to include additional averaging over stochastic variables. In this way, the assumption of time--translational invariance will remain justified throughout in spite of frictional forces.

In Sec. \ref{Theoretical Considerations}, we present the basic definitions and the correlation functions to be studied. In Sec. \ref{Static correlations}, the one--and two--particle densites and the static structure factor are obtained. In Sec. \ref{Dynamic correlations}, the intermediate scattering function has been calculated in the presence of damping including a stochastic force with Gaussian noise. Resulting low--friction density--relaxation spectra are discussed as function of temperature.  In Sec. \ref{Debye--Waller factor},  density excitations for the limiting case of zero friction and the Debye--Waller factor (DWF) are discussed. Finally, concluding remarks are given in Sec. \ref{Concluding remarks}.

\section{Theoretical Considerations}
\label{Theoretical Considerations}

In the description of dynamical properties of a many--particle system,  the correlation function of density fluctuations $\delta N({\bf r},t)$,
\1
\label{ddcf}
F({\bf r},{\bf r'},t)=\frac{1}{{\cal N}}\erw{\delta N({\bf r},t)\,\delta N({\bf
r'},0)}\;,
\2
and its spatial Fourier transform
\1
\label{ft-ddcf}
F_{{\bf q}\,{\bf q'}}(t)=\int\de^3r\int\de^3r'~\e^{-\ii {\bf
q}\cdot{\bf r}+\ii{\bf q'}\cdot{\bf r'}}F({\bf r},{\bf r'},t)=
\frac{1}{{\cal N}}\erw{\delta N_{\bf q}(t)\,\delta N_{\bf q'}(0)^*}
\2
play an important role. Here  $\erw{\dots}$ denotes a thermal equilibrium average, and the dynamical variable
$
N({\bf r},t)\g\sum_{j\g1}^{\cal N} \delta({\bf r}-{\bf r}_j(t))
$
describes the particle density at space point ${\bf r}$ and time $t$, which, when
integrated over all space, sums up to the total number of particles,
${\cal N}\g\int\de^3r~N({\bf r},t)$. Correspondingly, the density variable in Fourier space is given by
\1
\label{delta-N_q}
N_{\bf q}(t)=\int\de^3r~\e^{-\ii{\bf q}\cdot{\bf r}}\,N({\bf
r},t)=\sum_{j\g1}^{\cal N}\e^{-\ii{\bf q}\cdot{\bf r}_j(t)}
\2
with fluctuation $\delta N_{\bf q}(t)$. The thermal fluctuation of a dynamical variable $A(t)$ is defined as $\delta A(t)\equiv A(t)-\erw{A(t)}\g A(t)-\erw{A}$, where invariance of the thermal equilibrium state with respect to time translations renders $\erw{A(t)}$ independent of time $t$.

The diagonal elements $F({\bf q},t)\g F_{{\bf q}\,{\bf q}}(t)$
are known as intermediate scattering function, because they determine the double--differential coherent--scattering cross section observed when scattering neutrons or photons from the system. The cross section is proportional to $\int_{-\infty}^\infty\de t~\exp(\ii t\omega) \erw{N_{\bf q}(t)N_{\bf q}(0)^*}$. It is conventionally written as
\cite{ege:94}
\1
\label{conv-ddscs}
\left(\dnd{\omega~\de\hat{\Omega}}{^2\sigma}\right)^{\rm coh}
=\frac{p_f}{p_i}\,a^2\left[\frac{|\erw{N_{\bf q}}|^2}{\cal
N}\,\delta(\omega)+S({\bf q},\omega)\right]
\2
by introducing the van Hove function
\1
\label{def-vanHove}
S({\bf q},\omega)=\frac{1}{2\pi}\int_{-\infty}^\infty\de t~\e^{\ii t\omega}
F({\bf q},t)\;.
\2
In the coherent--scattering cross section Eq.(\ref{conv-ddscs}),
$\hbar{\bf q}\g{\bf p}_i-{\bf p}_f$ and $\hbar\omega\g E_i-E_f$ denote the momentum and
the energy, respectively, which is transferred in a scattering process from the projectile (neutron, e.g.) to the target system (gas), and $a$ denotes the coherent scattering length of the gas.

We note that the right--hand side of Eq.(\ref{conv-ddscs}) will represent a decomposition into {\em elastic} and {\em inelastic} scattering contributions if, and only if, $F({\bf q},\infty)\g0$ implying an ergodic many--particle system. For non--ergodic systems with $F({\bf q},\infty)\!>\!0$, the van Hove function $S({\bf q},\omega)$ itself will have an elastic peak of strength $F({\bf q},\infty)$ in addition to inelastic ($\omega\!\ne\!0$) contributions. As will be shown below, the harmonically trapped ideal gas represents not only an inhomogeneous but also a non--ergodic many--particle system.

In general, the elastic coherent--scattering intensity is measured by the DWF defined as the ratio of {\em elastic} and {\em total} coherent--scattering intensity,
\ea
\label{def-DWF}
f({\bf q})&=&\lim_{t\to0}\frac{\erw{N_{\bf q}(t)N_{\bf q}^*}}{\erw{|N_{\bf q}|^2}}=\frac{|\erw{N_{\bf
 q}}|^2/{\cal N}+F({\bf q},\infty)}{|\erw{N_{\bf
 q}}|^2/{\cal N}+F({\bf q},0)}\;.
\ee
Here
\1
\label{def-sofq}
F({\bf q},0)=\int_{-\infty}^\infty\de\omega~S({\bf q},\omega)\equiv S({\bf q})=\frac{1}{\cal N}\erw{|\delta N_{\bf q}|^2}
\2
is the well known static structure factor $S({\bf q})$. We emphasize that no assumptions regarding translational invariance of the target system have been made in the above equations.
We note that in Eq.(\ref{def-DWF}) there are two contributions to the DWF, in general. It is worth recalling that for a cristalline solid the first contribution is nonzero, while the second, i.e. $F({\bf q},\infty)$, vanishes since the cristalline solid is an {\em ergodic} system. On the other hand, for a non--ergodic (structural) glass $F({\bf q},\infty)>0$, and $\erw{N_{\bf  q}}$ is zero for all ${\bf q}\ne0$ in the homogeneous system. In a homogeneous fluid, both contributions are zero resulting in a vanishing DWF. However, in the inhomogeneous trapped gas studied here, there will be a finite DWF as shown in Sec. \ref{Debye--Waller factor} below.

Although $F_{{\bf q}\,{\bf q'}}(t)$ for an inhomogeneous system will generally be non--zero if ${\bf q}\!\ne\!{\bf q'}$, only the {\em diagonal} elements are required for calculating the coherent double--differential scattering cross section. That is, the intermediate scattering function $F({\bf q},t)\g F_{{\bf q}\,{\bf q}}(t)$ will determine the coherent--scattering cross section (aside from the average density, which is  responsible for the elastic `Bragg--scattering' contribution of strength $|\erw{N_{\bf
 q}}|^2/{\cal N}$).

We now consider a system of ${\cal N}$ non--interacting particles of
mass $m$ moving in three--dimensional space under the influence of
an isotropic harmonic external potential. This collection of oscillators is
described by the Hamiltonian
\1
\label{hamiltonian}
H=\sum_{j\g1}^{\cal N}\left[\frac{{\bf p}_j^2}{2m}+V({\bf
r}_j)\right]\;,
\2
where $V({\bf r})\g\frac{1}{2}m\Omega^2{\bf r}^2$
with oscillator frequency $\Omega$, and ${\bf p}_j$ and ${\bf r}_j$
denote momentum and position variables of the $j$--th particle,
respectively.
For the trapped ideal gas described by Eq.(\ref{hamiltonian}),  thermal averages of additive single--particle variables of the
general form  $A(t):=\sum_{j\g1}^{\cal N}A({\bf p}_j(t),{\bf
r}_j(t))$ reduce to the simple expression
\1
\label{average}
\erw{A(t)}={\cal N}\int\de^3\op{p}~\frac{ \exp\left( -\frac{\op{{\bf p}}^2}{2p_{\rm
T}^2} \right) }{\left(p_{\rm T}\sqrt{2\pi}\right)^3}\int\de^3\op{r}~\frac{\exp\left(-\frac{\op{{\bf
r}}^2}{2u^2}\right)}{ \left(u\sqrt{2\pi}\right)^3}
 ~A(\op{{\bf p}}(t),\op{{\bf r}}(t))\;,
\2
where $p_{\rm T}\g m v_T$, $v_T\g\sqrt{k_{\rm B}T/m}$, and
$u\g v_T/\Omega$ denote, respectively, the particle's thermal
momentum, thermal speed, and  distance traveled within time
$\Omega^{-1}$.
In general, $\op{{\bf r}}(t)\g {\bf R}(\op{{\bf p}},\op{{\bf r}};t)$ and $\op{{\bf p}}(t)\g {\bf P}(\op{{\bf p}},\op{{\bf r}};t)$ will depend on initial momentum $\op{{\bf p}}$, initial position $\op{{\bf r}}$, and time $t$. The vector--valued functions ${\bf R}$ and ${\bf P}$ will be determined from Hamilton's equations of motion in Sec.\ref{Dynamic correlations} below.
For $t\g0$, however, $\op{{\bf r}}(0)\g\op{{\bf r}}$ and $\op{{\bf p}}(0)\g\op{{\bf p}}$. The evaluation of the 6--fold integral on the right--hand side of Eq.(\ref{average}) is straight forward then. The static averages entering Eqs.(\ref{conv-ddscs}, \ref{def-DWF}, \ref{def-sofq}) are discussed in Sec.\ref{Static correlations}.

\section{Static correlations}
\label{Static correlations}

Using Eq.(\ref{average}), one obtains for the
average particle--number density,
\1
\label{av-density}
\erw{N({\bf r})}=\frac{{\cal
N}\exp\left(-\frac{{\bf
r}^2}{2u^2}\right)}{\left(u\sqrt{2\pi}\right)^3}~~\longleftrightarrow~~
\erw{N_{{\bf q}}}={\cal N}\e^{-\frac{1}{2}\,q^2u^2}\;.
\2
We also note that the mean--square displacement of a particle from the center of the trap is
\1
\label{msd}
\erw{{\bf r}_0^2}=\erw{\frac{1}{{\cal
N}}\sum_{j\g1}^{\cal N}{\bf r}_j^2}=3u^2\;,
\2
rendering an additional physical interpretation of the length $u$.
According to Eqs.(\ref{av-density}--\ref{msd}), in thermal equilibrium the trapped gas will form a spherically symmetric cloud in which 90.0(99.0; 99.9)\% of the gas particles are contained in a sphere of radius $a\!\approx\! 2.5 (3.4; 4.0)u$ around the center of force.

For the initial value $F_{{\bf q}\,{\bf q'}}(0)$ describing the static correlations of density fluctuations, one finds (see also Eqs.(\ref{fqqsvont-idgas}--\ref{csqt-explicit}) below):
\1
\label{static-fqqs}
F_{{\bf q}\,{\bf q'}}(0)=\frac{1}{{\cal N}}\erw{\delta N_{\bf q}\,\delta N_{\bf q'}^*}
=\e^{-\frac{1}{2}(q^2+q'^2)u^2}\left(\e^{{\bf q}\cdot{\bf q'}\,u^2}-1\right)\;.
\2
The corresponding result in ${\bf r}$--space is given in App.\ref{non--diagonal correlation functions}.
Note the symmetry relation, $F_{{\bf q}\,{\bf q'}}(0)\g F_{{\bf q'}\,{\bf q}}(0)$, and the vanishing small--$q$ limit, $\lim_{q\to0}F_{{\bf q}\,{\bf q'}}(0)\g0$.

At this stage, let us point out an interesting fact regarding the  static structure factor  defined in Eq.(\ref{def-sofq}) which, for the trapped ideal gas, can be read from the diagonal elements of Eq.(\ref{static-fqqs}):
\1
\label{sofq-idgas}
S({\bf q})=1-\e^{-q^2u^2}\;.
\2
As displayed in Fig. \ref{Fig-1}(a), $S(q)\to q^2u^2+{\cal O}(q^4)$ for long wavelengths.
Lacking a well--defined volume occupied by the ${\cal N}$ particles, an overall particle--number density $n$ cannot rigourously be defined in the inhomogeneous gas. However, the ``peak density''
\1
\label{overall-density}
n_0:=\erw{N({\bf r}\g0)}=\frac{{\cal N}}{L^3}\;,~~~~~~~~~~~~L:=u\sqrt{2\pi}
\2
presents a reasonable measure of the overall density. Due to the rapidly decreasing average density $\erw{N({\bf r})}$ for $|{\bf r}|>u$ (see Eq.(\ref{av-density})), $V\g L^3$ will be considered to represent the gas volume. It corresponds to a sphere of radius $a\g3^{1/3}(\pi/2)^{1/6}\approx1.56\,u$ containing about 51\% of the total number ${\cal N}$ of particles.
An estimate of $u$ for $^{87}$Rb in a trap of $\Omega\g 120\times 2\pi\,{\rm s^{-1}}$ at $T\g5\,T_c\approx735\,{\rm n K}$
gives $u\approx11\,{\rm \mu m}$. Assuming the trap to contain ${\cal N}\g2\times10^4$ atoms, the  peak density will be $n_0\approx 10^{12}\,{\rm cm^{-3}}$ which is typical of the values encountered in cold--atom traps.

For the non--uniform ideal gas considered here, one finds  for the average two--particle density \cite[Chap. 2.5]{ham:05}
\1
\rho_{\cal N}^{(2)}({\bf r},{\bf r}')=\frac{{\cal N}({\cal N}-1)}{\left(u\sqrt{2\pi}\right)^6}\,\exp\left(-\frac{{\bf r}^2+{\bf r}'^2}{2u^2}\right)\;,
\2
implying the two--particle distribution function
$
g_{\cal N}^{(2)}({\bf r},{\bf r}')=1-1/{\cal N}\;,
$
which is a constant depending on particle number ${\cal N}$.

\section{Dynamic correlations}
\label{Dynamic correlations}

It is straight--forward to show that the density--fluctuation correlation functions of the non--interacting gas take the simple form
\1
\label{fqqsvont-idgas}
F_{{\bf q}\,{\bf q'}}(t)=C_{{\bf q}\,{\bf q'}}^{(\rm s)}(t)-\e^{\frac{1}{2}(q^2+q'^2)u^2}
\2
with $C_{{\bf q}\,{\bf q'}}^{(\rm s)}(t)\g\frac{1}{\cal N}\sum_{j\g1}^{\cal N}\erw{\e^{-\ii{\bf q}\cdot{\bf r}_j(t)}\e^{\ii{\bf q'}\cdot{\bf r}_j(0)}}$ denoting the tagged--particle density correlation function, which reduces to
\ea
\label{csqt-explicit}
C_{{\bf q}\,{\bf q'}}^{(\rm s)}(t)&=&\int\de^3\op{p}~\frac{
\exp\left(-\frac{\op{{\bf p}}^2}{2p_{\rm T}^2}\right)
}{\left(p_{\rm T}\sqrt{2\pi}\right)^3}\int\de^3\op{r}~\frac{\exp\left(-\frac{\op{{\bf r}}^2}{2u^2}\right)}{
\left(u\sqrt{2\pi}\right)^3}
 ~\left\{\e^{-\ii{\bf q}\cdot\op{{\bf r}}(t)}\,\e^{\ii{\bf q'}\cdot\op{{\bf r}}}\right\}_{\rm av}\;,
\ee
where $\{\dots\}_{\rm av}$ explicitly indicates averaging over the stochastic--force variable.
Inserting into Eq.(\ref{csqt-explicit}) the solution $\op{{\bf r}}(t)$ according to Eq.(\ref{elongation}) with initial time $t_0\g0$ , we obtain
\ea
\label{correct-csqt}
C_{\rm s}(q,t)&=&C_{{\bf q}\,{\bf q}}^{(\rm s)}(t)=\exp\left\{-q^2u^2[1-\phi(t)]\right\}
\ee
with initial value $C_{\rm s}(q,0)\g1$ implied by Eq.(\ref{initval-phi}).
For non--zero friction, the long--time limiting value $C_{\rm s}(q,\infty)=\exp\left(-q^2u^2\right)$ follows from Eq.(\ref{ltl-phi}).
\begin{figure}
\begin{center}
\includegraphics[width=90mm,angle=0]{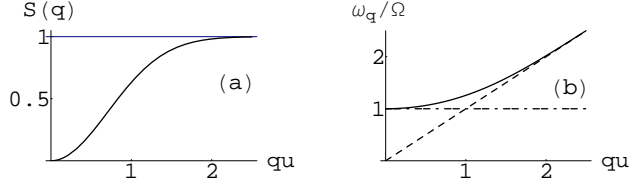}
\caption{\label{Fig-1}
(a) Static structure factor; (b) characteristic frequency for trapped gas (full line) and free gas (dashed), and $\omega\g\Omega$ (dash--dotted).
}
\end{center}
\end{figure}

\begin{figure}
\begin{center}
\includegraphics[width=90mm,angle=0]{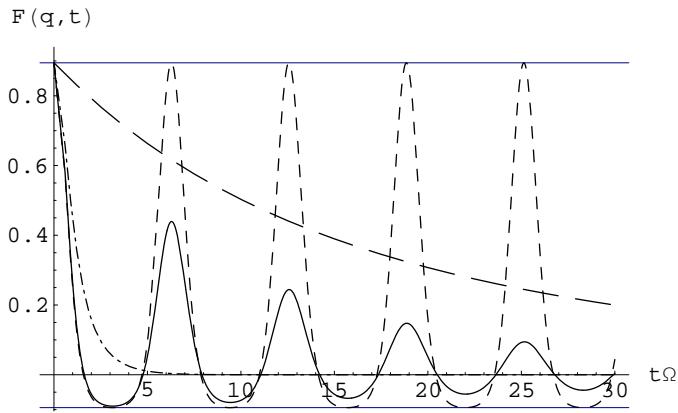}
\caption{\label{Fig-2}
Intermediate scattering functions at wavenumber $qu\g1.5$ for $\Gamma/\Omega\g0$ (dashed), 0.1 (full line), 2.01 (dash--dotted), and 40.0 (long--dashed). Horizontal lines indicate function range, Eq.(\ref{range-fsqt}), for this wavenumber.
}
\end{center}
\end{figure}

As a consequence, the intermediate scattering function becomes
\1
\label{def-fqt}
F(q,t)=F_{{\bf q}\,{\bf q}}(t)=\left[\e^{q^2u^2\phi(t)}-1\right]\e^{-q^2u^2}\;,
\2
which is displayed in Fig. \ref{Fig-2} for various $\Gamma$ and one representative wavenumber $q$,  will have a damped oscillatory form, in general, with amplitudes ranging between
\1
\label{range-fsqt}
-S(q)\,\e^{-q^2u^2}\le F(q,t)\le S(q)\;.
\2
It is readily seen from Eqs.(\ref{correct-csqt}),(\ref{ltl-phi}) that $\lim_{t\to\infty}F(q,t)\g0$ for any finite friction constant, and oscillations will be over--damped, if $\Gamma$ exceeds the  aperiodic--limit value $\Gamma\g2\Omega$.
Evidently, the intermediate scattering function of the trapped ideal gas is closely reflecting the motion of a single gas particle under the influence of friction and harmonic trap potential.

We note that the static structure factor $S(q)$ is not affected by the damping mechanism introduced above. Further, the short--time behavior,
\1
\label{Fofqt-exp}
F(q,t)=S(q)\left[1-\frac{1}{2}~\omega_{q}^2t^2+{\cal O}(t^3)\right]\;,~~~~~~~~
\omega_{q} =\frac{q v_T}{\sqrt{S(q)}}\;,
\2
is governed by $\omega_{q}$, the characteristic frequency defined by the second frequency moment $\omega_{q}^2:=\int\de\omega~\omega^2\phi''(q,\omega)$ of the normalized dynamic structure factor,
\1
\label{def-DFRS}
\phi''(q,\omega)=\pi \frac{S(q,\omega)}{S(q)}\;.
\2

The static structure factor, which is vanishing as $q^2$ for $q\to0$, is responsible for the gap of size $\Omega$ in the graph of $\omega_{q}$, see Fig. \ref{Fig-1}(b).
Physically, this frequency gap has the following meaning. Long--wavelength excitations of the trapped {\em ideal gas} do not constitute  propagating density fluctuations (phonons). Instead, a long--wavelength perturbation of the harmonically trapped gas will induce a collective oscillation with frequency $\Omega$ of the total gas mass.\cite{plasma:09}
The appearance of an excitation with frequency $\omega_q$ could be thought of as a hybridization of thermal free--particle excitations and the collective oscillation with frequency $\Omega$ associated with  the trap.
With increasing parameter $q u$, however, this mode will disappear merging into the continuous spectrum,
\1
\label{phi-free-id-gas}
\phi''(q,\omega)\stackrel{qu\gg1}{\longrightarrow}\frac{\pi}{S(q)}\,
\frac{\exp\left(-\frac{\omega^2}{2\omega_q^2}\right)}{\omega_q\sqrt{2\pi}}\,\;,
\2
which coincides with the well--known relaxation spectrum of density fluctuations of a uniform ideal gas for $S(q)\equiv1$.\cite{deriv:09}
As $q u$ is increased above one, the characteristic frequency will change its physical significance, because a rapidly growing number of additional resonances at $\omega\g0,\,\pm2\Omega,\,\pm3\Omega,\dots$ will contribute to the second--frequency moment (see Eq.(\ref{ssqom-class-undamped}) below). Eventually, for wavenumbers $qu\gg1$, the characteristic frequency $\omega_q\to qv_{ \rm T}$ will determine the thermal width of the broad Gaussian spectral line centered at zero frequency. Note that the normalized dynamic structure factor  has become independent of the trap frequency $\Omega$ in the free--particle limit  described by Eq.(\ref{phi-free-id-gas}).

\begin{figure}
\begin{center}
\includegraphics[width=90mm,angle=0]{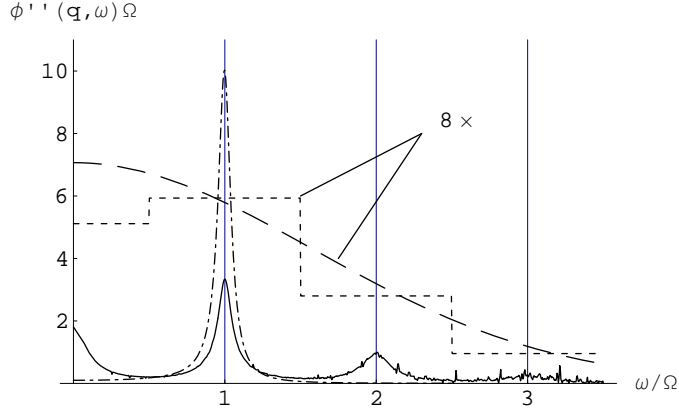}
\caption{\label{Fig-3}
Normalized dynamic structure factors at $T\g5 T_0$. For $\Gamma/\Omega\g0.1$, at $qu\g1.5$ (full line) from Eqs.(\ref{def-vanHove}),(\ref{def-DFRS}) and $q u\g0$ (dash--dotted) from Eq.(\ref{elong-relax-spec}). Enlarged by factor 8, for $\Gamma/\Omega\g0$, and at $qu\g1.5$: $\delta$--resonances (vertical grid lines) and weights (dashed stairway) from Eqs.(\ref{ssqom-class-undamped}),(\ref{def-DFRS}), and large--$(qu)$ asymptote (long--dashed) from Eq.(\ref{phi-free-id-gas}).
}
\end{center}
\end{figure}
\begin{figure}
\begin{center}
\includegraphics[width=90mm,angle=0]{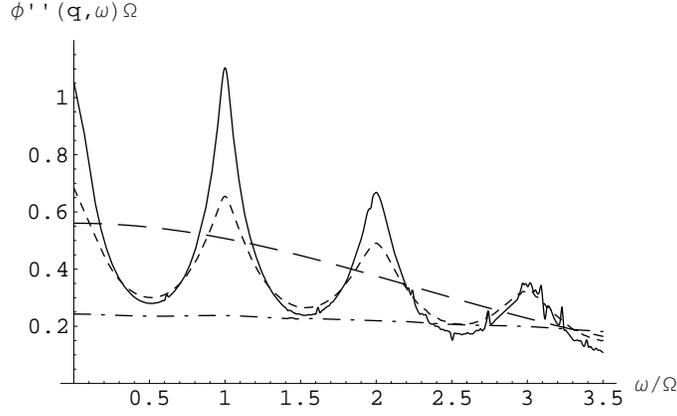}
\caption{\label{Fig-4}
Normalized dynamic structure factors for $\Gamma\g0.1\,\Omega$, $q\g1.5\,u_0^{-1}$, and $T/T_0\g20, \,7$ and 5 (dash--dotted, dashed, and full line, resp.) from Eqs.(\ref{def-vanHove}),(\ref{def-DFRS}), and for $T/T_0\g5$ (long--dashed) from Eq.(\ref{phi-free-id-gas}).
}
\end{center}
\end{figure}

Both $S(q,\omega)$ and $\phi''(q,\omega)$, Eq.(\ref{def-DFRS}), may be computed  from Eqs.(\ref{def-vanHove}),(\ref{def-sofq}),(\ref{correct-csqt}) for arbitrary $q>0$ and $\Gamma>0$. An example is given in Fig. \ref{Fig-3} showing $\phi''(q,\omega)$ (full line) at wavenumber $q u\g1.5$ of a trapped low--friction gas ($\Gamma\g 0.1\,\Omega$). It corresponds to the intermediate scattering function which is displayed in Fig. \ref{Fig-2} by a full line, too. Besides the friction--broadened oscillation peak at $\omega\g\Omega$, the spectrum clearly exhibits the above mentioned additional peaks appearing at $\omega\g0$ and $2\Omega$.
These additional peaks are absent in the corresponding spectrum taken at $qu\g0$, which is also displayed in Fig. \ref{Fig-3}.
In the long--wavelength limit, the normalized dynamic structure factor may be calculated analytically, since
\1
\label{elong-relax-spec}
\lim_{q\to0}\phi''(q,\omega)\equiv\phi''(\omega):=\frac{1}{2}\int_{-\infty}^\infty\de t~\e^{\ii t\omega}\,\phi(t)=\frac{\Gamma\Omega^2}{\left(\omega^2-\Omega^2\right)^2+\omega^2\Gamma^2}
\;,
\2
where $\phi''(\omega)$ denotes the relaxation spectrum of the oscillator elongation,
which reduces to a pair of $\delta$--functions for vanishing friction constant:
\1
\label{two-peaks}
\lim_{\Gamma\to0}\phi''(0,\omega)=\frac{\pi}{2}\left[\delta(\omega-\Omega)
+\delta(\omega+\Omega))\right]\;.
\2
For sufficiently small friction, $\phi''(0,\omega)$ describes the collective density excitation at frequency $\omega_{q\g0}\g\Omega$ exhibited in Fig. \ref{Fig-3} (dash--dotted curve). It should be noted, however, that its detection in a scattering experiment will be difficult due to the small overall scattering intensity $S(q)\approx (qu)^2$ associated with small values of the parameter $qu$. On the other hand, one has a large overall scattering intensity of $S(q)\approx 0.9$ for $qu\g1.5$, which corresponds to the full curve in Fig. \ref{Fig-3} and to all curves in Fig. \ref{Fig-4}.

At non--zero $q$ and sufficiently small $\Gamma$, one may speculate the spectrum $\phi''(q,\omega)$ of the form Eq.(\ref{elong-relax-spec}) but with $\Omega\to\Omega_q$ and $\Gamma\to \Gamma_q$, where $\Omega_q\approx \omega_q$ as $q\to0$.

Obviously, the damping mechanism introduced above cannot create positive dispersion ($\Omega_q>\Omega$).
Weak frictional and stochastic forces merely achieve broadening of the resonances at $\pm\Omega$ without shifting them. This broadening will only in the over--damped regime lead to an apparent negative dispersion associated with the collapse of peaks at $\omega\g0$ (see Eq.(\ref{elong-relax-spec})).

Figure \ref{Fig-4} displays variations of the normalized $S(q,\omega)$ induced by changing the temperature. If $u_0\g\sqrt{k_{\rm B}T_0/(m\Omega^2)}$ serves as unit of length, where $T_0$ denotes a reference temperature,  all $T$--dependence of the above formulas will be contained in $qu\g qu_0\sqrt{T/T_0}$.
Choosing a fixed wavenumber $q\g1.5\,u_0^{-1}$ and the BEC critical temperature of ${\cal N}\g2\times10^4$ $^{87}$Rb atoms in a trap with $\Omega\g2\pi\times120\,{\rm Hz}$ as a reference, i.e. $T_0\g T_c\approx147\,{\rm nK}$, the full curve in Fig. \ref{Fig-4} will correspond to $T\g5\,T_c\approx735\,{\rm nK}$, a temperature  well above the condensation point,  where the linear extension of the cloud is $L\approx28\,{\rm \mu m}$.

We note that, for a gas at an ultra--low temperature well above $T_c$, the difference between density--fluctuation spectra of a trapped and a free ideal gas is not only qualitative but quantitative as the comparison of full and long-dashed curves in Fig. \ref{Fig-4} may demonstrate. This makes us believe that the predicted spectra could be observed in experiments.

Although we chose a small damping constant $\Gamma\g0.1\Omega$ in producing Figs.\ref{Fig-3} and \ref{Fig-4}, the friction may be unrealistically large, if the results are interpreted as describing an ultra--cold gas cloud. Instead of producing similar plots from Eqs.(\ref{def-vanHove}),(\ref{def-DFRS}) for much smaller friction constant, which would introduce an increasing amount of difficulties in handling the numerical integration, we will present the analytical solution of the frictionless case in the following section.

\section{Undamped motion and DWF}
\label{Debye--Waller factor}

For a purely harmonic trap, one has $\phi(t)\g\cos(\Omega t)$ according to Eq.(\ref{oscillatorrelaxationfunction}). Using this in Eq.(\ref{def-fqt}), one obtains the intermediate scattering function
\ea
\label{fqt-no-friction}
F(q,t)&=&\e^{-q^2u^2}\left\{\exp\left[q^2u^2\cos(\Omega t)\right]-1\right\}\\
\label{fqt-harmonic}
&=&\e^{-q^2u^2}\left\{I_0(q^2u^2)-1+2\sum_{n\g1}^\infty I_n(q^2u^2)\cos(n\Omega t)\right\}\;,
\ee
where the generating function of modified Bessel functions of the first kind \cite[Chap. 9.6.34 ]{abs:70} was applied in the last line.
\begin{figure}
\begin{center}
\includegraphics[width=90mm,angle=0]{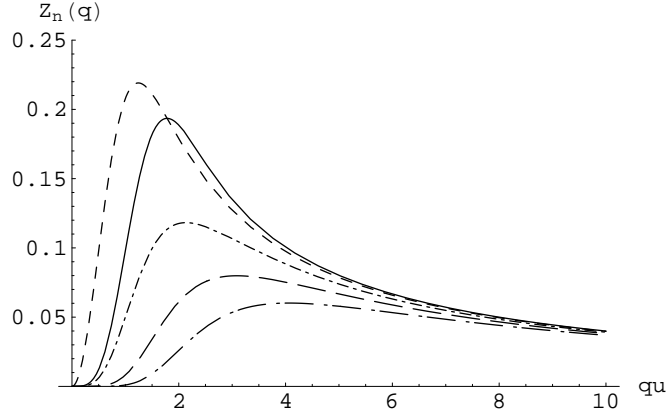}
\caption{\label{Fig-5}
Spectral weights of $S(q,\omega)$ for $n\g0$ (full), 1 (dashed), 2 (dash--dotted), 3 (long-dashed), 4 (long-dash--dotted).
}
\end{center}
\end{figure}
Noting $I_{-n}(z)\g I_{n}(z)$, the dynamic structure factor resulting from Eq.(\ref{fqt-harmonic})
reads explicitly:
\1
\label{ssqom-class-undamped}
S(q,\omega)=Z_0(q)\,\delta(\omega)+\sum_{n\g1}^\infty Z_n(q)
\left[\delta(\omega-n \Omega)+\delta(\omega+n \Omega)\right]\;.
\2
with spectral weights
\1
Z_n(q):=\e^{-q^2u^2}\left[I_n(q^2u^2)-\delta_{n,0}\right]=Z_{-n}(q)
\2
corresponding to excitation peaks situated at $\omega\g n \Omega$ ($n\g0,\,\pm1,\,\pm2,\dots$), which sum up to $\sum_{n\g-\infty}^\infty Z_n(q)= S(q)$.

Weights are displayed in Fig. \ref{Fig-5} as function of the dimensionless parameter $qu$. The weights of the elastic and the first inelastic peak at $\omega\g\pm\Omega$ have pronounced maxima, while all other $Z_n(q)$ also show a maximum of smaller magnitude around $qu\g2$,   decrease to zero slowly for $q\to\infty$, and vanish at $q\g0$. For $|n|\ge1$, they decrease with increasing $n$.
Further, it can be seen that $Z_1$ clearly dominates the inelastic--scattering weights  $Z_n$ in the range of small wavenumbers, $1>qu>0$.
In the long--wavelength limit, finally, the collection of infinitely many resonances will collapse into a pair at $\omega\g\pm\Omega$, in accordance with Eq.(\ref{two-peaks}). That is easily verified by applying the limiting values
\1
\lim_{q\to0}\frac{Z_n(q)}{S(q)}=\frac{1}{2}\,\delta_{|n|,1}\;,
\2
in Eq.(\ref{ssqom-class-undamped}) from which one finds
\1
S(q,\omega)\longrightarrow q^2u^2/2\left[\delta(\omega-\Omega)+\delta(\omega+\Omega)\right]
\2
for $qu\ll1$.
Spectral weights $\pi Z_n(q)/S(q)$ corresponding to $\phi''(q,\omega)$ have been indicated  for $qu\g1.5$ in Fig. \ref{Fig-3} in terms of  a `stairway' spectral function together with the continuous spectrum Eq.(\ref{phi-free-id-gas}).  Plotting a sequence of such stairways for increasing values of the parameter $qu$ will provide a good visualization of the spectrum metamorphosis from a sum of only a few visible stairs for $qu<1$ to the continuous broad spectrum of the uniform ideal gas for $qu\gg1$.

\begin{figure}
\begin{center}
\includegraphics[width=90mm,angle=0]{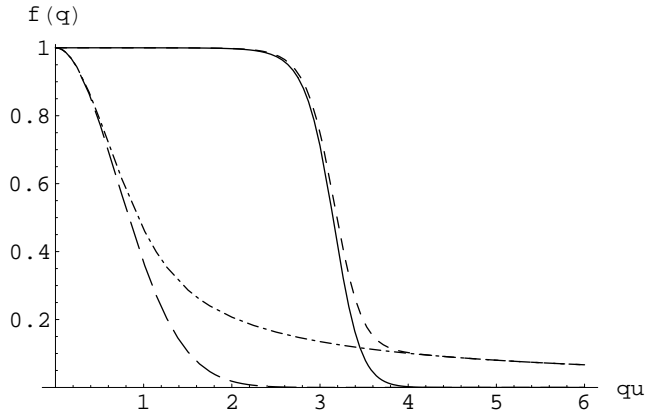}
\caption{\label{Fig-6}
Debye--Waller factor of ${\cal N}\g 2.\times 10^4$ atoms in harmonic trap for $\Gamma>0$ (full line) and $\Gamma\g0$ (dashed). DWF of single trapped atom (${\cal N}\g1$) for $\Gamma>0$ (long--dashed) and $\Gamma\g0$ (dash--dotted).
}
\end{center}
\end{figure}

It is to be noted that for the purely harmonic trap the  dynamical
structure factor $S(q,\omega)$ defined by the temporal Fourier
transform of the correlation function of density fluctuations,
Eq.(\ref{def-vanHove}), will contain an elastic scattering
contribution due to $F(q,\infty)\ne0$, which is absent for
$\Gamma>0$. This contribution, together with the a--priori elastic
contribution $\delta(\omega)|\erw{N_{\bf q}}|^2/{\cal N}$ to the
coherent double--differential scattering cross section, when used in Eq.(\ref{def-DWF}), will lead to DWF of the trapped ideal gas
\1
\label{trapped-id-gas-DWF}
f(q)=\frac{\e^{-q^2u^2}\left[I_0(q^2u^2)+{\cal N}-1\right]}{1+({\cal
N}-1)\e^{-q^2u^2}}\;, \2 whereas for $\Gamma>0$ the bracket in the
numerator of Eq.(\ref{trapped-id-gas-DWF}) is simply replaced with
${\cal N}$. Contrary to a homogeneous gas, the DWF of the trapped
atomic cloud does not vanish and depends on ${\cal N}$. The DWF is
displayed for a trapped cloud of ${\cal N}\g20,000$ atoms in
Fig.\ref{Fig-6}. For comparison, also the DWF of a single trapped atom
(${\cal N}\g1$) is given, which coincides with the
Lamb--M{\"o}{\ss}bauer factor (LMF) of the trapped ideal gas.

\section{Concluding remarks}
\label{Concluding remarks}

In this paper we have calculated the density--fluctuation spectrum of a classical ideal gas in a harmonic potential at temperatures well above the BEC temperature, but in a micro Kelvin range which is still of interest. In this regime, $\hbar\Omega$ is much less than $k_{\rm B}T$ for trapped atomic gases. It is found that the difference between the density fluctuation spectra of the trapped and the free gas is large enough to be observed experimentally. The calculation predicts the existence of density excitations of the trapped mass which increases with increasing wave number. It is shown that Debye--Waller factor for trapped thermal clouds of ${\cal N}$ atoms are finite contrary to a homogeneous gas, opening the possibility of observing recoilless emission of ultra--low frequency radiation from the trapped gas. It is hoped that this work will stimulate interest in performing inelastic scattering experiments.

\vspace{2cm}
{\large\bf Acknowledgements}\\*[3mm]
KNP and JB acknowledge financial support from Alexander-von-Humboldt Foundation, Germany, and KNP acknowledges the University Grants Commission, New Delhi, for financial support in the form of emeritus fellowship.
The work is partially supported by the grant for visits under Indo-German (DST-DFG) collaborative research program.

\section*{Appendix}

\renewcommand{\theequation}{\mbox{\Alph{section}.\arabic{equation}}}
\begin{appendix}
\section{Time evolution of $\op{\bf r}(t)$ and $\op{\bf p}(t)$}
\label{Time evolution}
\setcounter{equation}{0}

In order to evaluate Eq.(\ref{csqt-explicit}), we need to know the time evolution of particle position
$\op{{\bf r}}(t)\g {\bf R}(\op{{\bf p}},\op{{\bf r}},t)$ and momentum $\op{{\bf p}}(t)\g {\bf P}(\op{{\bf p}},\op{{\bf r}},t)$. In the present case of an isotropic oscillator potential, the vector components of $\op{\bf r}(t)$ as well as of $\op{\bf p}(t)$ are not coupled. Therefore we outline the method of evaluation of the $x$--components $\op{x}(t)$ and $\op{p}_x(t)$, only. The same method applies to the $y$-- and $z$--components.

In the presence of a driving force ${\cal  F}_x(t)$, the elongation $\op{x}(t)$ will obey the damped driven  oscillator--equations of motion,
\ea
\label{osc-eom1}
\der{t}\op{x}(t)&=&\frac{1}{m}\,\op{p}_x(t)\\
\label{osc-eom2}
\der{t}\op{p}_x(t)&=&-m\Omega^2~\op{x}(t)-m\Gamma~\der{t}\op{x}(t)+{\cal F}_x(t)\;,
\ee
with the unique solution
\1
\label{elongation}
\op{x}(t)=\phi(t-t_0)\, \op{x}+\ii\chi(t-t_0)\,\op{p}_x+\int_{t_0}^t\de t'~\ii\chi(t-t')\,{\cal F}_x(t')
\2
corresponding to initial conditions $\op{x}(t_0)\g\op{x}$ and $\op{p}_x(t_0)\g\op{p}_x$.
Here $\phi(t)$ and $\chi(t)$, defined by ($t\ge 0$)
\ea
\label{oscillatorrelaxationfunction}
\phi(t)&:=&\frac{z_1 e^{z_2 t}-z_2 e^{z_1 t}}{z_1-z_2}\;,~~~~~~~~\chi(t):=\frac{1}{\ii m}\,\frac{\e^{z_1 t}-\e^{z_2 t}}{z_1-z_2}
\;,\\
\label{rootsofcharacteristicpolynomial}
z_{1,2}&=&-\Gamma/2\pm\sqrt{(\Gamma/2)^2-\Omega^2}\;,
\ee
denote the (normalized) {\em relaxation function} and the {\em linear response function} of the oscillator elongation, respectively. Equations similar to Eq.(\ref{elongation}) also apply for the $y$-- and $z$--components.

Indeed, $\phi(t)$ represents the normalized Kubo relaxation function of elongation. This can be seen as follows. Choosing $t_0\g0$ in Eq.(\ref{elongation}), multiplying with $\op{x}/(k_{\rm B}T)$ on both sides and subsequently taking a thermal average will ---on the left--hand side--- result in the Kubo relaxation function of elongation, $\Phi(t)\g\erw{\op{x}(t)\op{x}}/(k_{\rm B}T)$. On the right--hand side, only the first term $\phi(t)\,\erw{\op{x}^2}/(k_{\rm B}T)\equiv\phi(t)\Phi(0)$ will survive, where $\Phi(0)\g1/(m\Omega^2)$.

Both functions, $\phi(t)$ and $\chi(t)$, are related by Kubo's identity,
\1
\label{KI}
\chi(t)=\frac{\ii\dot{\phi}(t)}{m\Omega^2}\;,
\2
which can be verified explicitly for the results given in Eq.(\ref{oscillatorrelaxationfunction}). Also, the initial values of the relaxation function,
\1
\label{initval-phi}
\phi(0)=1\;,~~~~~~~~~~\dot{\phi}(0)=0\;,
\2
can be read from Eqs.(\ref{oscillatorrelaxationfunction}),(\ref{KI}), while one has
\1
\label{ltl-phi}
\phi(\infty)=0\;,~~~~~~~~~~~~\dot{\phi}(\infty)=0\;,~~~~~~~~~~~~(\Gamma>0)
\2
for finite friction.
The magnitude of the normalized relaxation function is restricted to the interval $[0,1]$ or, equivalently,
\1
-1\le\phi(t)\le1
\2
for any time $t$.

In addition to the restoring force and the friction force, it has been assumed in Eq.(\ref{osc-eom2}) that there acts a time--dependent random force ${\cal F}_x(t)$ on the oscillator.
Thermal averaging is then understood to include additional averaging over stochastic variables as indicated explicitly in Eq.(\ref{csqt-explicit}). For evaluating the stochastic average, we assumed that the driving force ${\cal F}_x(t)$  represents Gaussian white noise, which after straightforward calculation leads to the result Eq.(\ref{correct-csqt}).

\section{Non--diagonal correlation functions}
\label{non--diagonal correlation functions}
\setcounter{equation}{0}
Assuming undamped oscillations
\1
\op{{\bf r}}(t)=\cos(\Omega t)\,\op{\bf r}+\frac{\sin(\Omega t)}{m\Omega}\,\op{\bf p}\;,
\2
one finds from Eqs.(\ref{fqqsvont-idgas}--\ref{csqt-explicit})
\1
\label{fqqst}
F_{{\bf q}\,{\bf q'}}(t)=\e^{-\frac{1}{2}(q^2+q'^2)u^2}\left[\e^{{\bf q}\cdot{\bf q'}\,u^2\cos(\Omega t)}-1\right]\;,
\2
which by Fourier transformation leads to
\1
\label{frrst}
F({\bf r},{\bf r'},t)=\left[\frac{\exp\left(-\frac{\left[{\bf
r}-{\bf r'}\cos(\Omega t)\right]^2}{2u^2\sin^2(\Omega t)}\right)}{\left(u|\sin(\Omega t)|\sqrt{2\pi}\right)^3}-\frac{\exp\left(-\frac{{\bf
r}^2}{2u^2}\right)}{\left(u\sqrt{2\pi}\right)^3}\right]\frac{\exp\left(-\frac{{\bf
r'}^2}{2u^2}\right)}{\left(u\sqrt{2\pi}\right)^3}\;.
\2
It is noted that $F_{{\bf q}\,{\bf q'}}(t)\g F_{{\bf q'}\,{\bf q}}(t)$ and $\lim_{q\to0}F_{{\bf q}\,{\bf q'}}(t)\g0$. We further observe
$\int\de^3r~F({\bf r},{\bf r'},t)\g\int\de^3r'~F({\bf r},{\bf r'},t)\g0$.

Taking the limit $t\to0$ in Eq.(\ref{frrst}) one finds
\1
\label{static-frrs}
F({\bf r},{\bf r'},0)=\left[\delta({\bf r}-{\bf r'})-\frac{\exp\left(-\frac{{\bf
r}^2}{2u^2}\right)}{\left(u\sqrt{2\pi}\right)^3}\right]\frac{\exp\left(-\frac{{\bf
r'}^2}{2u^2}\right)}{\left(u\sqrt{2\pi}\right)^3}\;.
\2
which is the inverse Fourier transform of Eq.(\ref{static-fqqs}) or of the $t\g0$--value of Eq.(\ref{fqqst}).

\end{appendix}

\bibliographystyle{unsrt}

\begin{thebibliography}{10}
\bibitem{plasma:09}
This is analogous to charge--density excitations in a one--component plasma
  where, in the long--wavelength limit, the charge density will perform
  oscillations with the plasma frequency due to restoring Coulomb forces
  created by the charge--neutralizing background.(see, e.g. \cite{bok.a:78}).
\bibitem{deriv:09}
A derivation of Eq.(\ref{phi-free-id-gas}) in the present context is
  straightforward. Starting from $\left.F(q,t)\right|_{\Gamma\g0}$ given in
  Eq.(\ref{fqt-no-friction}), inserting $u\g v_T/\Omega$ and $\cos(\Omega
  t)\to1-\Omega^2t^2/2$ for small $\Omega$, and performing the free--particle
  limit $\Omega\to0$, immediately results in $F(q,t)\g\exp(-q^2v_T^2t^2/2)$ for
  a free ideal gas which implies Eq.(\ref{phi-free-id-gas}).
\end{thebibliography}
\end{document}